\documentclass[thmsa,a4paper,oneside,onecolumn,12pt,final,titlepage]{article}
\usepackage{sw20lart}

\input{tcilatex}
\begin{document}

\title{Complete Correlation, Detection loophole and Bell's Theorem}
\author{A. Shafiee\thanks{%
E-mail: shafiee@sharif.edu}$^{*}$ \ and \ M. Golshani$^{\dagger }\bigskip $%
\quad \\
%EndAName
{\small \ }$\stackrel{*}{}$ {\small Department of Chemistry, Sharif
University of Technology,}\\
{\small \ P.O.Box 11365-9516, Tehran, Iran}\\
{\small \ }$\stackrel{\dagger }{}$ {\small Department of Physics, Sharif
University of Technology,}\\
{\small \ P.O.Box 11365-9161, Tehran, Iran.}}
\maketitle

\begin{abstract}
Two new formulations of Bell's theorem are given here. First, we consider a
definite set of two entangled photons with only two polarization directions,
for which Bell's locality assumption is violated for the case of perfect
correlation. Then, using a different approach, we prove an efficient
Bell-type inequality which is violated by some quantum mechanical
predictions, independent of the efficiency factors.
\end{abstract}

\section{Introduction}

Since its inception, Bell's theorem [1] has gone through many conceptual and
physical developments. However, even now, there are important aspects of
this theorem, which still demand more deliberation. Here, we are going to
focus on two such topics with both conceptual and practical significance.

This paper has two different themes. The first theme is about a
demonstration of Bell's theorem for the case of complete correlation, when
the two remote analyzers are along the same direction for a given pair of
entangled particles. For the case of two entangled photons\footnote{%
Hereafter, we always consider the photonic version of Bell's theorem in
which a source emits pairs of entangled photons with parallel linear
polarizations.}, we are dealing with a situation in which for different
sub-ensembles, the polarizations of each entangled photon are measured
simultaneously along different directions. For example, to show the
violation of Clauser-Holt-Shimony-Horne (CHSH) inequality [2], one should
measure the correlation functions $C(\widehat{a},\widehat{b})$, $C(\widehat{%
a^{\prime }},\widehat{b})$, $C(\widehat{a},\widehat{b^{\prime }})$ and $C(%
\widehat{a^{\prime }},\widehat{b^{\prime }})$ along four pairs of directions 
$(\widehat{a},\widehat{b})$, $(\widehat{a^{\prime }},\widehat{b})$, $(%
\widehat{a},\widehat{b^{\prime }})$ and $(\widehat{a^{\prime }},\widehat{%
b^{\prime }})$, respectively. At the same time, we should assume a certain
definition of \textit{non-contextuality} along with \textit{locality} in a
hidden-variable theory where the polarization of each photon along a
definite direction is supposed to be the same in all different contexts,
including different polarization directions of the remote analyzers.
However, one may propose a contextual local hidden-variable theory in which
the value of each observable comes about as the effect of the
system-apparatus local interaction [3]. Furthermore, it is possible to
assume local common causes which may not be the same in different contexts.
Now, the question arises as to \textit{why} (in a conceptual sense, not
merely algebraically) we cannot use a definite sub-ensemble of particle
pairs having only two polarization directions, to prove the impossibility of
getting quantum mechanical results by using local realism. And even going
further, would it be possible to show any inconsistency for the case of
perfect correlation? This situation leads to a stronger version of Bell's
theorem which its consequences should be scrutinized.

On the other hand, it is now apparent that there has not been any
experimental confirmation of the Bell experiments without having a loophole
[4]. As a result, people have attempted to derive more efficient
inequalities (for example see [5]) or have performed new experiments to
overcome the detection loophole [6]. This is the main theme of the second
part of the present paper. In the photonic experiments, the most difficulty
is due to the inefficiency of detectors which makes it impossible to measure
the polarization states of all photon pairs. This is the so-called
detection-efficiency loophole which P. Grangier describes as ``Achill's heel
of experimental tests of Bell's inequalities'' [7]. Consequently, some
auxiliary assumptions are introduced for demonstrating the experimental
violation of Bell's inequality. Among them, the most famous one is the fair
sampling assumption which means that undetected photons do not change the
statistical results of the experiment [8]. Nevertheless, this loophole is
still with us and Bell experiments are still being done.

Based on these considerations, in the first part of the paper, we prove a
new version of Bell's theorem for a definite sub-ensemble of correlated
photons. Then, we show that a stochastic local hidden-variable (SLHV) theory
cannot reproduce the quantum mechanical predictions for the case of \textit{%
perfect correlation}. The fact that the inconsistency is present in the
state of complete correlation has special significance.

In the second part of the paper, we provide more details about a new
inequality (similar to the Clauser-Horne (CH) inequality [9]) which has been
recently suggested as a short note [10]. We show that in a real photonic
experiment, if we consider the inefficiency of the measuring instruments,
our proposed inequality is violated by quantum mechanical results, \textit{%
independent} of the inefficiencies. This makes a good ground for doing
photonic experiments with more conclusive results.

\section{Bell's Theorem for The Case of Complete Correlation}

Consider an ideal Bell photonic experiment in which the entangled photons
are described by the state $\mid \psi \rangle =\frac{1}{\sqrt{2}}\left( \mid
H\rangle _{1}\mid H\rangle _{2}+\mid V\rangle _{1}\mid V\rangle _{2}\right) $
where, e.g., $\mid H\rangle _{1}$ ($\mid V\rangle _{1}$) stands for the
horizontal (vertical) polarization state of the first photon. Thus, the
entangled photons have parallel linear polarizations. In such experiment, if
the first photon passes through a filter with orientation $\widehat{a}$, its
mate crosses its corresponding filter in the same direction.

Now, consider a SLHV theory, where $\lambda $ represent all hidden variables
which belong to a space $\Lambda $. At the level of hidden variables, the
probability of registering the result $r$ for the polarization of the first
photon along $\widehat{a}$ is represented by $p_{r}^{(1)}(\widehat{a}%
,\lambda )$, where $r$ is equal to $+1$ or $-1$, depending on whether it is
passing through its filter or not. Similarly, the probability of registering
the value $q$ for the polarization measurement of the second photon along $%
\widehat{b}$ is represented by $p_{q}^{(2)}(\widehat{b},\lambda )$. The
average value of the outcomes of the polarization of the photon $k$ ($k=1$, $%
2$) along $\widehat{u}_{k}$ is given by:

\begin{equation}
\epsilon ^{(k)}(\widehat{u}_{k},\lambda )=\stackunder{j=\pm 1}{\dsum }j\
p_{j}^{(k)}(\widehat{u}_{k},\lambda )  \tag{1}
\end{equation}
where $\widehat{u}_{1}=\widehat{a}$ and $\widehat{u}_{2}=\widehat{b}$. From
this relation, it is clear that

\begin{equation}
\mid \epsilon ^{(k)}(\widehat{u}_{k},\lambda )\mid \leq 1  \tag{2}
\end{equation}

Using Bell's locality assumption, the average value of the outcomes of
polarizations of two photons along $\widehat{a}$ and $\widehat{b}$ is:

\begin{eqnarray}
\epsilon ^{(12)}(\widehat{a},\widehat{b},\lambda ) &=&\stackunder{r,q=\pm 1}{%
\dsum }rq\ p_{r}^{(1)}(\widehat{a},\lambda )\ p_{q}^{(2)}(\widehat{b}%
,\lambda )  \nonumber \\
&=&\epsilon ^{(1)}(\widehat{a},\lambda )\epsilon ^{(2)}(\widehat{b},\lambda )
\tag{3}
\end{eqnarray}
This relation originates from the fact that we have assumed there is
statistical independence at the level of hidden variables and that any
common characteristic of the two photons in the past is represented by
common causes $\lambda $. Correspondingly, the relation between the
correlation of the two polarizations at the experimental level, with the
average of the product of two polarizations at the hidden-variable level, is
given by:

\begin{equation}
C(\widehat{a},\widehat{b})=\dint\nolimits_{\Lambda }\epsilon ^{(12)}(%
\widehat{a},\widehat{b},\lambda )\ \rho (\lambda )d\lambda  \tag{4}
\end{equation}
where $\rho (\lambda )$ is the normalized distribution of the hidden
variables $\lambda $. From the relation (4), one can conclude that

\begin{equation}
\mid C(\widehat{a},\widehat{b})\mid \leq \dint\nolimits_{\Lambda }\mid
\epsilon ^{(12)}(\widehat{a},\widehat{b},\lambda )\ \mid \rho (\lambda
)d\lambda  \tag{5}
\end{equation}
But, $\mid \epsilon ^{(12)}(\widehat{a},\widehat{b},\lambda )\ \mid \leq 1$.
Thus,

\begin{equation}
\mid C(\widehat{a},\widehat{b})\mid \leq \dint\nolimits_{\Lambda }\mid
\epsilon ^{(12)}(\widehat{a},\widehat{b},\lambda )\ \mid \rho (\lambda
)d\lambda \leq 1  \tag{6}
\end{equation}

Now, consider an ideal experiment in which $\widehat{a}=\widehat{b}$. In
this case, we have a perfect correlation and one expects that $\mid C(%
\widehat{a},\widehat{a})\mid $ to be equal to one. Then, the inequality in
(6) turns into the following equality:

\begin{equation}
\dint\nolimits_{\Lambda }\left[ \mid \epsilon ^{(12)}(\widehat{a},\widehat{a}%
,\lambda )\ \mid -1\right] \rho (\lambda )d\lambda =0  \tag{7}
\end{equation}
which means that $\mid \epsilon ^{(12)}(\widehat{a},\widehat{a},\lambda )\
\mid $ should be equal to one. But, assuming Bell's locality condition in
(3), one gets:

\begin{equation}
\mid \epsilon ^{(1)}(\widehat{a},\lambda )\mid \mid \epsilon ^{(2)}(\widehat{%
a},\lambda )\mid =1\qquad \forall \ \widehat{a},\lambda  \tag{8}
\end{equation}
Thus, for \textit{any} $\widehat{a}$ \textit{and} $\lambda $ we must have $%
\mid \epsilon ^{(1)}(\widehat{a},\lambda )\mid =1$ which is a restrictive
form of the relation (2). This is in contradiction with the general
definition of a SLHV Theory.

Also, the context dependence of the hidden variables is irrelevant here. The
assumption of contextuality is always introduced when there are three or
more observables. Here, only two observables are involved: An observable is
assigned to the polarization of the first photon along $\widehat{a}$ and the
other observable to the polarization of the second photon along the same
direction. In the usual formulation of Bell's theorem, a third observable
could be assigned, e.g., to the polarization of the first photon along $%
\widehat{a^{\prime }}\neq \widehat{a}$ in a different measuring setup. So,
one has to consider different contexts of measuring setups, and whether the
value attributed to a quantity in different contexts is the same or not.
But, in this approach, the polarization of each photon is measured only in a
single experimental context and the assumption of contextuality cannot be
the reason behind the indicated contradiction.

One could simply show that the same conclusion is achieved by considering
any of the four entangled Bell states for both photonic and spin half
particles\footnote{%
For example, for spin half particles, the four Bell states are usually
denoted by $\mid \psi ^{\pm }\rangle =\frac{1}{\sqrt{2}}\left( \mid
z+\rangle _{1}\mid z-\rangle _{2}\pm \mid z-\rangle _{1}\mid z+\rangle
_{2}\right) $ and $\mid \phi ^{\pm }\rangle =\frac{1}{\sqrt{2}}\left( \mid
z+\rangle _{1}\mid z+\rangle _{2}\pm \mid z-\rangle _{1}\mid z-\rangle
_{2}\right) $ where, e.g., $\mid z+\rangle _{1}$ ($\mid z-\rangle _{1}$) is
the spin up (spin down) state of particle 1 along $z$-direction.}. Thus, we
can conclude that:

\begin{quote}
\textit{In a SLHV theory, Bell's locality condition does not coherently hold
for a definite measuring context of pairs of entangled particles, described
by any of the four Bell states, even for the case of complete correlation. }
\end{quote}

This is a new version of Bell's theorem for the state of complete
correlation, which we call hereafter as BTCC. It is, however, important to
notice that the relation (8) sounds legitimate in a deterministic local
hidden-variable (DLHV) theory. To see the reason, we define $A(\widehat{a}%
,\lambda )=\pm 1$ and $B(\widehat{b},\lambda )=\pm 1$ as the first and the
second photons' polarizations along $\widehat{a}$ and $\widehat{b}$,
respectively, in a DLHV theory. Subsequently, one can write Bell's locality
condition as a similar relation to (3). This means that $\epsilon
_{DET}^{(12)}(\widehat{a},\widehat{b},\lambda )=A(\widehat{a},\lambda )B(%
\widehat{b},\lambda )$ where $\epsilon ^{(1)}(\widehat{a},\lambda )$ ($%
\epsilon ^{(2)}(\widehat{b},\lambda )$) is replaced by $A(\widehat{a}%
,\lambda )$ ($B(\widehat{b},\lambda )$) in (3) and $\left| \epsilon
_{DET}^{(12)}(\widehat{a},\widehat{b},\lambda )\right| =1$ [11]. Then, if
one follows the same discussion as above for a DLHV theory, one will reach
the conclusion that $\mid A(\widehat{a},\lambda )\mid $ should be equal to
one. This result is trivial, however, because $A(\widehat{a},\lambda )$
accepts only the values $+1$ and $-1$. Thus, BTCC does not include DLHV
theories.

Even for those local hidden-variable theories in which $\lambda $ uniquely
determine the polarization state of \textit{one} of the particles (e.g., if
only for the first particle we have $\epsilon ^{(1)}(\widehat{a},\lambda )=A(%
\widehat{a},\lambda )=\pm 1$), BTCC cannot be concluded. Any appearance of
the deterministic behavior in $\lambda $ for either of the particles would
break down the above argument. So, it seems that a fundamental discrepancy
between the stochastic and the deterministic local hidden-variables theories
exists which reveals itself when the complete correlation is taken into
account: The DLHV theories are more \textit{robust} to fall into
contradiction than the SLHV ones. This is in contrast to the common belief
that the stochastic nature of hidden variables makes the scope of Bell's
theorem much broader, because determinism seems to be a particular case of a
probabilistic formulation (at least, relationally), when each probability
reaches the determinate values of zero and one.

There is also another reason which strengthens this claim: No Bell photonic
experiment performed so far can prove the incongruity of the DLHV theories.
It is because the auxiliary assumptions imposed to see the violation of
Bell's inequalities have \textit{statistical} character and their meaning is
obscure in a deterministic approach. For example, you may consider the fair
sampling assumption when it is used in CHSH inequality. This assumption
means that unrecorded data do not have a weighty role in calculating the
polarization correlations of the entangled photons. More precisely, this
means that there are alternative effective correlation functions such as

\begin{equation}
C_{eff}(\widehat{a},\widehat{b})=\dint\nolimits_{\Lambda }\epsilon
_{eff}^{(12)}(\widehat{a},\widehat{b},\lambda )\ \rho (\lambda )d\lambda 
\tag{9}
\end{equation}
which are similarly bounded by CHSH inequality whenever the \textit{hidden}
average value $\epsilon _{eff}^{(12)}(\widehat{a},\widehat{b},\lambda )$ as
well as the other similar expressions along different directions satisfy the
CHSH inequality [12]. (The value of $\epsilon _{eff}^{(12)}(\widehat{a},%
\widehat{b},\lambda )$ for a given $\lambda $ is determined by the kind of
the hidden-variable model which is used.) Therefore, according to the fair
sampling assumption, we should have:

\begin{equation}
\left| C_{eff}(\widehat{a},\widehat{b})+C_{eff}(\widehat{a^{\prime }},%
\widehat{b})+C_{eff}(\widehat{a^{\prime }},\widehat{b^{\prime }})-C_{eff}(%
\widehat{a},\widehat{b^{\prime }})\right| \leq 2  \tag{10}
\end{equation}
where $C_{eff}(\widehat{a},\widehat{b})$, e.g., is defined as: 
\begin{equation}
C_{eff}(\widehat{a},\widehat{b})=\frac{N_{++}^{(12)}(\widehat{a},\widehat{b}%
)+N_{--}^{(12)}(\widehat{a},\widehat{b})-N_{-+}^{(12)}(\widehat{a},\widehat{b%
})-N_{+-}^{(12)}(\widehat{a},\widehat{b})}{N_{++}^{(12)}(\widehat{a},%
\widehat{b})+N_{--}^{(12)}(\widehat{a},\widehat{b})+N_{-+}^{(12)}(\widehat{a}%
,\widehat{b})+N_{+-}^{(12)}(\widehat{a},\widehat{b})}  \tag{11}
\end{equation}
in terms of the number of joint detections $N_{rq}^{(12)}(\widehat{a},%
\widehat{b})$ for the results $r,q=\pm 1$. A convenient way for realizing of
the performance of the fair sampling assumption is as follows. Using the
predictions of a SLHV model, we should first obtain a relation for $\epsilon
^{(12)}(\widehat{a},\widehat{b},\lambda ),$ which includes the non-detection
results too. Then, the value of $\epsilon ^{(12)}(\widehat{a},\widehat{b}%
,\lambda )$ should be renormalized on the basis of the detected results.
This is $\epsilon _{eff}^{(12)}(\widehat{a},\widehat{b},\lambda )$.
Subsequently, we should check to see if the inequality (10) can be obtained
on the basis of relation (9) where the effective correlation functions in
(10) are defined as (11). Of course, it is a well known fact that many SLHV
models have been suggested which could deny the assumption of fair sampling
(for example see [13]).

Similarly, one may define $\epsilon _{eff,DET}^{(12)}(\widehat{a},\widehat{b}%
,\lambda )=A_{eff}(\widehat{a},\lambda )B_{eff}(\widehat{b},\lambda )$ for a
DLHV theory. But, what is the meaning of the \textit{efficient} determinate
value $A_{eff}(\widehat{a},\lambda )$ (and/or $B_{eff}(\widehat{b},\lambda )$%
)? If an efficient determinate value means that only detected results should
be involved, its meaning improperly reduces to the case of an ideal
experiment. On the other side, there is no alternative definition consistent
with what we expect to mean for $C_{eff}(\widehat{a},\widehat{b})$ at the
experimental level. Hence, the meaning of the fair sampling assumption is
completely vague in DLHV theories.

In summary, for two reasons, the formulation of BTCC provides more evidence
for the insufficiency of a SLHV theory. The first reason is that such a
formulation excludes any context dependence of the stochastic hidden
variables, because the context is the same for all the photon pairs. It
presents a stronger inconsistency of the SLHV theories, also because it
includes the case of complete (anti)correlation. Secondly, it reveals a
fundamental discrepancy between the SLHV and the DLHV theories, which is
concealed in the usual formulations of Bell's theorem.

Nevertheless, the problem of doing an \textit{actual} Bell experiment is
still with us. If one can prove a more efficient inequality (as a type of CH
or CHSH inequalities) in which no auxiliary assumption is used at the level
of hidden-variables, its violation in real experiments can weaken the
possibility of both the SLHV and the DLHV theories. This is the main theme
of the next section\footnote{%
It should be emphasized, however, that any derivation of the Bell
inequalities based on the original formulation of Bell's theorem ignores the
possibility of contextual hidden variables, because it should be always
assumed that in different measuring setups, the same hidden variables should
be considered.}.

\section{An Extended CH Inequality}

When one faces the detection loophole in Bell experiments, a natural
question may arise as to why the violation of the Bell inequalities are so
sensitive to the efficiency factors in real examinations. This subject is so
important that nowadays some people believe that there may be some (unknown)
physical constraints in nature which could prevent us from doing perfect
experiments [14]. Nevertheless, since the beginning of quantum mechanics,
there were many peculiar features of the quantum particles (e.g., the
wave-particle duality of massive particles, tunneling effect and so on)
which have been confirmed in experiments, even with not perfect instruments.
Hence, what is specific in the \textit{correlation} of two spatially
separated particles which makes us so skeptical? If there is nothing
special, there should be a way to find a loophole-free Bell-type inequality
which could be easily violated in experiments.

To find a solution for this old problem, let us consider a Bell photonic
experiment in which the linear polarizations of the photon are measured
along $(\widehat{a},\widehat{b})$, $(\widehat{a},\widehat{b^{\prime }})$, $(%
\widehat{a^{\prime }},\widehat{b})$, and $(\widehat{a^{\prime }},\widehat{%
b^{\prime }})$. Now, we define a function $g_{rq}$ as:

\begin{eqnarray}
g_{rq}(\widehat{a},\widehat{b},\widehat{a^{\prime }},\widehat{b^{\prime }}%
,\lambda ) &=&p_{r}^{(1)}(\widehat{a},\lambda )\ \left[ p_{q}^{(2)}(\widehat{%
b},\lambda )-p_{q}^{(2)}(\widehat{b^{\prime }},\lambda )\right]  \nonumber \\
&&+p_{r}^{(1)}(\widehat{a^{\prime }},\lambda )\ \left[ p_{q}^{(2)}(\widehat{b%
},\lambda )+p_{q}^{(2)}(\widehat{b^{\prime }},\lambda )\right]  \nonumber \\
&&-p_{r}^{(1)}(\widehat{a^{\prime }},\lambda )\ p_{r}^{(2)}(\widehat{%
a^{\prime }},\lambda )-p_{q}^{(1)}(\widehat{b},\lambda )\ p_{q}^{(2)}(%
\widehat{b},\lambda )  \tag{12}
\end{eqnarray}
where the probability functions $p_{r}^{(1)}$ and $p_{q}^{(2)}$ ($r,q=\pm 1$%
) are defined as before. Since, $g_{rq}$ is a linear function of the
single-particle probabilities $p_{r}^{(1)}(\widehat{a},\lambda )$, $%
p_{q}^{(2)}(\widehat{b},\lambda )$, $p_{r}^{(1)}(\widehat{a^{\prime }}%
,\lambda )$, $p_{q}^{(2)}(\widehat{b^{\prime }},\lambda )$, $p_{r}^{(2)}(%
\widehat{a^{\prime }},\lambda )$ and $p_{q}^{(1)}(\widehat{b},\lambda )$,
its upper and lower bounds are determined by the limits of these variables.
Now, in an ideal experiment, if we consider all sixteen possible
combinations of zero and one for the functions $p_{r}^{(1)}(\widehat{a}%
,\lambda )$, $p_{q}^{(2)}(\widehat{b},\lambda )$, $p_{r}^{(1)}(\widehat{%
a^{\prime }},\lambda )$ and $p_{q}^{(2)}(\widehat{b^{\prime }},\lambda )$,
when $p_{r}^{(1)}(\widehat{a^{\prime }},\lambda )$ is zero or one, $%
p_{r}^{(2)}(\widehat{a^{\prime }},\lambda )$ \textit{should be} also zero or
one and the same holds for $p_{q}^{(2)}(\widehat{b},\lambda )$ and $%
p_{q}^{(1)}(\widehat{b},\lambda )$. This is due to the fact that the photon
pairs have parallel polarizations. Thus, if the first photon passes through
an analyzer with the polarization direction $\widehat{a^{\prime }}$ (i.e., $%
p_{+}^{(1)}(\widehat{a^{\prime }},\lambda )=1$), the second photon also
passes through an analyzer with the same direction of polarization (i.e., $%
p_{+}^{(2)}(\widehat{a^{\prime }},\lambda )=1$) and vice versa. In an actual
experiment, however, the situation is a little different: The upper limit of
relation (12) is not necessarily equal to zero in non ideal experiments. To
explain this fact, let us define the sum of the detection probabilities as:

\begin{equation}
\stackunder{j=\pm 1}{\dsum }\ p_{j}^{(k)}(\widehat{x}_{k},\lambda )=\alpha
^{(k)}(\widehat{x}_{k},\lambda )=1-p_{0}^{(k)}(\widehat{x}_{k},\lambda ) 
\tag{13}
\end{equation}
where $\widehat{x}_{1}=\widehat{a}$, $\widehat{a^{\prime }}$ or $\widehat{b}$%
, $\widehat{x}_{2}=\widehat{b}$, $\widehat{b^{\prime }}$ or $\widehat{%
a^{\prime }}$ and $k=1,\ 2$. The function $p_{0}^{(k)}(\widehat{x}%
_{k},\lambda )$ denotes a non-detection probability for the $k$th photon
with the polarization along $\widehat{x}_{k}.$ So, $\alpha ^{(k)}(\widehat{x}%
_{k},\lambda )$ is a measure of inefficiencies at the level of
hidden-variables. Table 1 shows the possible values of the lower and upper
limits of the single-particle probabilities in terms of the inefficiency
measures defined in (13). For more convenience, we have denoted $\alpha
^{(1)}(\widehat{a},\lambda )\equiv \alpha _{1}$, $\alpha ^{(1)}(\widehat{%
a^{\prime }},\lambda )\equiv \alpha _{1}^{\prime }$, $\alpha ^{(2)}(\widehat{%
a^{\prime }},\lambda )\equiv \alpha _{2}^{\prime }$, $\alpha ^{(2)}(\widehat{%
b},\lambda )\equiv \beta _{2},$ $\alpha ^{(2)}(\widehat{b^{\prime }},\lambda
)\equiv \beta _{2}^{\prime }$ and $\alpha ^{(1)}(\widehat{b},\lambda )\equiv
\beta _{1}$. We exclude here the possibilities of either $p_{r}^{(1)}(%
\widehat{a^{\prime }},\lambda )$ ($p_{q}^{(2)}(\widehat{b},\lambda )$) or $%
p_{r}^{(2)}(\widehat{a^{\prime }},\lambda )$ ($p_{q}^{(1)}(\widehat{b}%
,\lambda )$) being zero but not both. This situation is in conflict with the
assumption of parallel polarization, if one takes into account the ideal
limits of table 1.\bigskip The lower and upper limits of $g_{rq}$ are then
classified in table 2.

$
\begin{tabular}{|c|c|c|c|c|c|c|}
\hline
Rows & $p_{r}^{(1)}(\widehat{a},\lambda )$ & $p_{q}^{(2)}(\widehat{b}%
,\lambda )$ & $p_{r}^{(1)}(\widehat{a^{\prime }},\lambda )$ & $p_{q}^{(2)}(%
\widehat{b^{\prime }},\lambda )$ & $p_{r}^{(2)}(\widehat{a^{\prime }}%
,\lambda )$ & $p_{q}^{(1)}(\widehat{b},\lambda )$ \\ \hline
1 & $0$ & $0$ & $0$ & $0$ & $0$ & $0$ \\ \hline
2 & $\alpha _{1}$ & $0$ & $0$ & $0$ & $0$ & $0$ \\ \hline
3 & $0$ & $\beta _{2}$ & $0$ & $0$ & $0$ & $\beta _{1}$ \\ \hline
4 & $0$ & $0$ & $\alpha _{1}^{\prime }$ & $0$ & $\alpha _{2}^{\prime }$ & $0$
\\ \hline
5 & $0$ & $0$ & $0$ & $\beta _{2}^{\prime }$ & $0$ & $0$ \\ \hline
6 & $0$ & $0$ & $\alpha _{1}^{\prime }$ & $\beta _{2}^{\prime }$ & $\alpha
_{2}^{\prime }$ & $0$ \\ \hline
7 & $0$ & $\beta _{2}$ & $0$ & $\beta _{2}^{\prime }$ & $0$ & $\beta _{1}$
\\ \hline
8 & $\alpha _{1}$ & $0$ & $0$ & $\beta _{2}^{\prime }$ & $0$ & $0$ \\ \hline
9 & $0$ & $\beta _{2}$ & $\alpha _{1}^{\prime }$ & $0$ & $\alpha
_{2}^{\prime }$ & $\beta _{1}$ \\ \hline
10 & $\alpha _{1}$ & $0$ & $\alpha _{1}^{\prime }$ & $0$ & $\alpha
_{2}^{\prime }$ & $0$ \\ \hline
11 & $\alpha _{1}$ & $\beta _{2}$ & $0$ & $0$ & $0$ & $\beta _{1}$ \\ \hline
12 & $0$ & $\beta _{2}$ & $\alpha _{1}^{\prime }$ & $\beta _{2}^{\prime }$ & 
$\alpha _{2}^{\prime }$ & $\beta _{1}$ \\ \hline
13 & $\alpha _{1}$ & $0$ & $\alpha _{1}^{\prime }$ & $\beta _{2}^{\prime }$
& $\alpha _{2}^{\prime }$ & $0$ \\ \hline
14 & $\alpha _{1}$ & $\beta _{2}$ & $0$ & $\beta _{2}^{\prime }$ & $0$ & $%
\beta _{1}$ \\ \hline
15 & $\alpha _{1}$ & $\beta _{2}$ & $\alpha _{1}^{\prime }$ & $0$ & $\alpha
_{2}^{\prime }$ & $\beta _{1}$ \\ \hline
16 & $\alpha _{1}$ & $\beta _{2}$ & $\alpha _{1}^{\prime }$ & $\beta
_{2}^{\prime }$ & $\alpha _{2}^{\prime }$ & $\beta _{1}$ \\ \hline
\end{tabular}
$

\bigskip \textsl{Table 1: The possible values of single-particle
probabilities }

\qquad \qquad \textsl{in an actual experiment.\bigskip }

$
\begin{tabular}{|c|c|}
\hline
Rows & $g_{rq}$ \\ \hline
1 & $0$ \\ \hline
2 & $0$ \\ \hline
3 & $-\beta _{2}\beta _{1}$ \\ \hline
4 & $-\alpha _{1}^{\prime }\alpha _{2}^{\prime }$ \\ \hline
5 & $0$ \\ \hline
6 & $\alpha _{1}^{\prime }(\beta _{2}^{\prime }-\alpha _{2}^{\prime })$ \\ 
\hline
7 & $-\beta _{2}\beta _{1}$ \\ \hline
8 & $-\alpha _{1}\beta _{2}^{\prime }$ \\ \hline
9 & $\alpha _{1}^{\prime }(\beta _{2}-\alpha _{2}^{\prime })-\beta _{2}\beta
_{1}$ \\ \hline
10 & $-\alpha _{1}^{\prime }\alpha _{2}^{\prime }$ \\ \hline
11 & $(\alpha _{1}-\beta _{1})\beta _{2}$ \\ \hline
12 & $\alpha _{1}^{\prime }(\beta _{2}^{\prime }-\alpha _{2}^{\prime
})+(\alpha _{1}^{\prime }-\beta _{1})\beta _{2}$ \\ \hline
13 & $(-\alpha _{1}+\alpha _{1}^{\prime })\beta _{2}^{\prime }-\alpha
_{1}^{\prime }\alpha _{2}^{\prime }$ \\ \hline
14 & $\alpha _{1}(\beta _{2}-\beta _{2}^{\prime })-\beta _{2}\beta _{1}$ \\ 
\hline
15 & $(\alpha _{1}-\beta _{1})\beta _{2}+\alpha _{1}^{\prime }(\beta
_{2}-\alpha _{2}^{\prime })$ \\ \hline
16 & $\left( \alpha _{1}-\beta _{1}\right) \beta _{2}+(\alpha _{1}^{\prime
}-\alpha _{1})\beta _{2}^{\prime }+\alpha _{1}^{\prime }(\beta _{2}-\alpha
_{2}^{\prime })$ \\ \hline
\end{tabular}
\bigskip $

\textsl{Table 2: The limits of }$g_{rq}$.\bigskip

It is now apparent that the upper limit of $g_{rq}$ can exceed zero in the
rows 6, 11, 12, 15 and 16 in table 2. But, it is possible to keep the zero
limit at the \textit{experimental} level as we explain below. Let us first
average the rows 6, 11, 12, 15 and 16 in table 2 over the space $\Lambda $.
According to the definition of the inefficiency measures in (13), for
example, we get for row 11:

\begin{eqnarray}
&&\int_{\Lambda }\left[ \alpha ^{(1)}(\widehat{a},\lambda )-\alpha ^{(1)}(%
\widehat{b},\lambda )\right] \alpha ^{(2)}(\widehat{b},\lambda )\rho
(\lambda )d\lambda  \nonumber \\
&=&\ \int_{\Lambda }\stackunder{r,q=\pm 1}{\dsum }\left[ p_{r}^{(1)}(%
\widehat{a},\lambda )-p_{r}^{(1)}(\widehat{b},\lambda )\right] p_{q}^{(2)}(%
\widehat{b},\lambda )\rho (\lambda )d\lambda  \nonumber \\
&=&\stackunder{r,q}{\sum }\left[ P_{rq}^{(12)}(\widehat{a},\widehat{b}%
)-P_{rq}^{(12)}(\widehat{b},\widehat{b})\right]  \tag{14}
\end{eqnarray}
where, e.g., $P_{rq}^{(12)}(\widehat{a},\widehat{b})$ is the joint
probability for getting the results $r$ and $q$ for the first and second
photons along $\widehat{a}$ and $\widehat{b}$, respectively in the
experiment, and $\rho (\lambda )$ is a probability density in space $\Lambda 
$. One can easily show that the right hand side of relation (14) is equal to 
$\left( 1-P_{0}^{(2)}(\widehat{b})\right) \left( P_{0}^{(1)}(\widehat{b}%
)-P_{0}^{(1)}(\widehat{a})\right) $, assuming that the joint probability of
non-detection is factorizable, i.e., $P_{00}^{(12)}=P_{0}^{(1)}P_{0}^{(2)}$.
It is now obvious that if one assumes that the non-detection probabilities $%
P_{0}^{(1)}(\widehat{b})$ and $P_{0}^{(1)}(\widehat{a})$ are equal, the
relation (14) will become zero. This means in turn that the non-detection
probabilities for the first particle are independent of the polarization
direction. The same calculations for rows 6, 12, 15 and 16 in table 2 will
show that if the non-detection probabilities for \textit{both} of the
particles are assumed to be independent of the polarization direction, the
average value of each of those rows will be equal to zero. For the other
rows in table 2, the same argument shows that the average values lie in the $%
\left[ -1,0\right] $ interval. So, one can obtain the following inequality:

\begin{equation}
-1\leq G_{rq}(\widehat{a},\widehat{b},\widehat{a^{\prime }},\widehat{%
b^{\prime }})\leq 0  \tag{15}
\end{equation}
where $G_{rq}(\widehat{a},\widehat{b},\widehat{a^{\prime }},\widehat{%
b^{\prime }})=\int\nolimits_{\Lambda }g_{rq}(\widehat{a},\widehat{b},%
\widehat{a^{\prime }},\widehat{b^{\prime }},\lambda )\rho (\lambda )d\lambda 
$ and $g_{rq}(\widehat{a},\widehat{b},\widehat{a^{\prime }},\widehat{%
b^{\prime }},\lambda )$ was defined in (12). Hence, $G_{rq}(\widehat{a},%
\widehat{b},\widehat{a^{\prime }},\widehat{b^{\prime }})$ is equal to:

\begin{eqnarray}
G_{rq}(\widehat{a},\widehat{b},\widehat{a^{\prime }},\widehat{b^{\prime }})
&=&P_{rq}^{(12)}(\widehat{a},\widehat{b})-P_{rq}^{(12)}(\widehat{a},\widehat{%
b^{\prime }})+P_{rq}^{(12)}(\widehat{a^{\prime }},\widehat{b})+P_{rq}^{(12)}(%
\widehat{a^{\prime }},\widehat{b^{\prime }})  \nonumber \\
&&-P_{rr}^{(12)}(\widehat{a^{\prime }},\widehat{a^{\prime }})-P_{qq}^{(12)}(%
\widehat{b},\widehat{b})  \tag{16}
\end{eqnarray}
where in deriving (16), we have used Bell's locality assumption. For
example, we have used relations such as:

\begin{equation}
P_{rq}^{(12)}(\widehat{a},\widehat{b})=\dint_{\Lambda }p_{r}^{(1)}(\widehat{a%
},\lambda )\ p_{q}^{(2)}(\widehat{b},\lambda )\ \rho (\lambda )d\lambda 
\tag{17}
\end{equation}

Evidently, in reaching the inequality (15) the following assumption is vital:

\begin{quote}
\textbf{A}- \textit{The experimental probabilities of non-detection for each
of the particles are independent of the polarization directions in each wing
of a Bell-type photonic experiment.}
\end{quote}

The validity of assumption \textbf{A} can be checked in the experiments.
This can be done by checking if the practical efficiency parameters change,
when one changes the polarization directions in analyzers in many trials of
a Bell experiment. This is the main difference between \textbf{A} and the
other auxiliary assumptions used before, because the assumption \textbf{A}
is \textit{testable}. It \textit{is not} introduced at the hidden-variable
level. For comparison, e.g., it is constructive to remember the CHSH
auxiliary assumption which is the same as \textbf{A} except for the term 
\textit{experimental }[2, 15].

The inequality (15) can be tested in non-ideal experiments. Here, we define $%
P_{rq}^{(12)}(\widehat{a},\widehat{b})$ as

\begin{equation}
P_{rq}^{(12)}(\widehat{a},\widehat{b})=\dfrac{N_{rq}^{(12)}(\widehat{a},%
\widehat{b})}{N_{tot}^{(12)}}  \tag{18}
\end{equation}
where $N_{rq}^{(12)}(\widehat{a},\widehat{b})$ is the number of photon pairs
whose polarization measurements, for photons 1 and 2 along $\widehat{a}$ and 
$\widehat{b}$ has yielded $r$ and $q$, respectively, and $N_{tot}^{(12)}$ is
the total number of photons emitted from the source. If we define other
probability functions in (16) similar to (18), after some algebra we get
from (15):

\begin{equation}
\dfrac{N_{rq}^{(12)}(\widehat{a},\widehat{b})-N_{rq}^{(12)}(\widehat{a},%
\widehat{b^{\prime }})+N_{rq}^{(12)}(\widehat{a^{\prime }},\widehat{b}%
)+N_{rq}^{(12)}(\widehat{a^{\prime }},\widehat{b^{\prime }})}{N_{rr}^{(12)}(%
\widehat{a^{\prime }},\widehat{a^{\prime }})+N_{qq}^{(12)}(\widehat{b},%
\widehat{b})}\leq 1  \tag{19}
\end{equation}
where $N_{tot}^{(12)}$ is eliminated. In this inequality we only deal with
recorded events. Of course, the CH inequality too, leads to a relation
similar to (19). But, the important thing is that in real experiments, where
we are dealing with inefficient detection, quantum mechanical predictions do
not violate the CH inequality [9, 15]. However, the inequality (15) can be
violated by quantum mechanical predictions. For example, in accordance with
the predictions of quantum mechanics, one may consider the joint probability 
$P_{rq,QM}^{(12)}(\widehat{a},\widehat{b})$ as:

\begin{equation}
P_{rq,QM}^{(12)}(\widehat{a},\widehat{b})\approx \frac{1}{4}\eta
_{overall}\left[ 1+\ rqF\cos 2(\widehat{a}-\widehat{b})\right]  \tag{20}
\end{equation}
where $\eta _{overall}$ is the overall efficiency of the devices (apart from
the efficiencies of the analyzers which are assumed to be approximately
perfect) and $F$ is a measure of the correlation of the two photons. Then,
by substituting (20) and relations similar to it in (16) and choosing $%
r=q=+1 $, we get

\begin{equation}
G_{++,QM}(\varphi )\approx \frac{1}{4}\eta _{overall}F\left[ 3\cos \varphi
-\cos 3\varphi -2\right]  \tag{21}
\end{equation}
where $\frac{\varphi }{2}=\mid \widehat{a}-\widehat{b}\mid =$ $\mid \widehat{%
a^{\prime }}-\widehat{b}\mid =$ $\mid \widehat{a^{\prime }}-\widehat{%
b^{\prime }}\mid $ and $\frac{3\varphi }{2}=\mid \widehat{a}-\widehat{%
b^{\prime }}\mid $. Considering the zero limit in the inequality (15), it is
then straightforward to show that the inequality can be violated for certain
ranges of $\varphi $ independent of the efficiency factors [10].

Our extended CH inequality is an instance of a set of inequalities in which
a similar argument can be used to show the inconsistency. For example, one
can define the following function instead of $g_{rq}$ in (12):

\begin{eqnarray}
f_{rq}(\widehat{a},\widehat{b},\widehat{a^{\prime }},\lambda )
&=&-p_{r}^{(1)}(\widehat{a},\lambda )\ p_{q}^{(2)}(\widehat{b},\lambda ) 
\nonumber \\
&&+p_{r}^{(1)}(\widehat{a^{\prime }},\lambda )\ \left[ p_{q}^{(2)}(\widehat{b%
},\lambda )+p_{r}^{(2)}(\widehat{a},\lambda )-p_{r}^{(2)}(\widehat{a^{\prime
}},\lambda )\right]  \tag{22}
\end{eqnarray}
Then, one can use the same assumption \textbf{A} to prove the following
inequality which can be tested in the experiments:

\begin{equation}
-1\leq -P_{rq}^{(12)}(\widehat{a},\widehat{b})+P_{rq}^{(12)}(\widehat{%
a^{\prime }},\widehat{b})+P_{rr}^{(12)}(\widehat{a^{\prime }},\widehat{a}%
)-P_{rr}^{(12)}(\widehat{a^{\prime }},\widehat{a^{\prime }})\leq 0  \tag{23}
\end{equation}

For some definite angles (e.g., $\mid \widehat{a^{\prime }}-\widehat{b}\mid
= $ $\mid \widehat{a}-\widehat{a^{\prime }}\mid =\frac{\theta }{2},\mid 
\widehat{a}-\widehat{b}\mid =\theta =\frac{\pi }{3}$), this is violated by
quantum mechanical predictions. However, it is an open problem whether one
can reduce the experimental settings (e.g., from three in (23) to two or
even one definite setting), while preserving the same limit of zero.

\section{Conclusion}

The meaning of ``entanglement'' and its empirical verification (as it
appears in Bell's theorem) demand still more elucidation. Here, we first
formulated a different version of Bell's theorem for two entangled photons
in the case of complete correlation (called BTCC). In the original
derivations of Bell's inequality, one has to introduce the local hidden
variables either counterfactually or noncontextually. According to BTCC,
however, the measuring context is the same for all the photon pairs because
it involves only the case of perfect (anti)correlation. Thus, BTCC
illustrates a stronger inconsistency of the SLHV theories. Afterwards, an
important discrepancy between the SLHV theories on one hand and the DLHV
theories on the other hand appears which is beyond the scope of Bell's
theorem. Hence, any demonstration of the inconsistency of the local
deterministic hidden variables calls for more strictness.

We also proved an extension of the CH inequality which indeed improves the
earlier efforts for demonstrating the violation of the Bell inequalities in
real experiments. The violation of our proposed inequality is independent of
the efficiency factors. But a crucial assumption is essential here which can
be tested in the experiments. This opens the door for a more reasonable
realization of the experimental results in the Bell photonic
experiments.\medskip

\textbf{Acknowledgment. }One of the authors (A. Shafiee) would like to thank
Apollo Go, Guillaume Adenier and Al Kracklauer for their valuable
discussions on an earlier version of this paper.

\end{document}